\documentclass[nofootinbib]{revtex4}
\usepackage{amsmath}
\usepackage{graphicx}
\usepackage{amssymb}
\usepackage{color}
\usepackage{amsfonts}

\begin{document}

\title{Weak gravitational lensing by compact objects in fourth order gravity}
\author{Zsolt Horv\'{a}th}
\email{zshorvath@titan.physx.u-szeged.hu}
\affiliation{Departments of Theoretical and Experimental Physics, University of Szeged, D%
\'{o}m t\'{e}r 9, Szeged 6720, Hungary}
\author{L\'{a}szl\'{o} \'{A}. Gergely}
\email{gergely@physx.u-szeged.hu}
\affiliation{Departments of Theoretical and Experimental Physics, University of Szeged, D%
\'{o}m t\'{e}r 9, Szeged 6720, Hungary}
\author{David Hobill}
\email{hobill@ucalgary.ca }
\affiliation{Department of Physics and Astronomy, University of Calgary, Calgary Alberta
T2N 1N4, Canada}
\author{Salvatore Capozziello}
\email{capozzie@na.infn.it}
\affiliation{Dipartimento di Scienze Fisiche, Universit`a di Napoli \textquotedblleft
Federico II\textquotedblright , I-80126, Napoli, Italy\\
INFN Sez. di Napoli, Compl. Univ. di Monte S. Angelo, Edificio G, Via
Cinthia, I-80126, Napoli, Italy}
\author{Mariafelicia De Laurentis}
\email{felicia@na.infn.it}
\affiliation{Dipartimento di Scienze Fisiche, Universit`a di Napoli \textquotedblleft
Federico II\textquotedblright, I-80126, Napoli, Italy\\
INFN Sez. di Napoli, Compl. Univ. di Monte S. Angelo, Edificio G, Via
Cinthia, I-80126, Napoli, Italy}

\begin{abstract}
We discuss weak lensing characteristics in the gravitational field of a
compact object in the low-energy approximation of fourth order $f\left(
R\right) $\ gravity theory. The particular solution is characterized by a
gravitational strength parameter $\sigma $\ and a distance scale $r_{c}$\
much larger than the Schwarzschild radius. Above $r_{c}$ gravity is
strengthened and as a consequence weak lensing features are modified
compared to the Schwarzschild case. We find a critical impact parameter
(depending upon $r_{c}$) for which the behavior of the deflection angle
changes. Using the Virbhadra-Ellis lens equation we improve the computation
of the image positions, Einstein ring radii, magnification factors and the
magnification ratio. We demonstrate that the magnification ratio as function
of image separation obeys a power-law depending on the parameter $\sigma $,
with a double degeneracy. No $\sigma \neq 0$ value gives the same power as
the one characterizing Schwarzschild black holes. As the magnification ratio
and the image separation are the lensing quantities most conveniently
determined by direct measurements, future lensing surveys will be able to
constrain the parameter $\sigma $ based on this prediction.
\end{abstract}

\maketitle

\section{Introduction}

The recent advent of the so-called \textquotedblleft Precision
Cosmology\textquotedblright\ along with galactic observations indicate that
General Relativity (GR) with standard matter sources disagrees with an
increasing number of observational data, e. g. those coming from IA-type
Supernovae, used as standard candles, large scale structure ranging from
galaxies up to superclusters \cite{SNeIa,CMBR,WMAP}, and galactic rotation
curves. In addition, from a theoretical point of view, being not
renormalizable, GR fails to be quantized in any \textit{standard} way (see 
\cite{uti}). Therefore at the extreme ultra-violet and infrared scales GR is
not and cannot be the definitive theory of Gravitation despite the fact that
it successfully addresses a wide range of phenomena and the Newtonian weak
field limit is correctly recovered.

In order to interpret the recent observational data in the framework of GR,
the introduction of unknown \textit{dark matter} (DM) (to address dynamical
phenomena as the formation of self-gravitating astrophysical structures) and 
\textit{dark energy} (DE) (to address the problem of cosmic acceleration)
seems to be necessary: however, the price of preserving the simplicity of
the Hilbert-Einstein Lagrangian has been the introduction of rather
odd-behaving physical entities that, up to now, have not been revealed by
any experiment at fundamental scales. This situation has led to several
attempts devoted either to recover the validity of GR at any scale, or to
construct alternative gravity theories that suitably generalize the
Einsteinian one. The philosophy of these two schemes is that in the former
case one has to modify the matter sector introducing DM and DE, in the
latter approach the dynamics of the geometry (i.e. the left-hand-side of the
Einstein equations) is modified but with the constraint to recover GR at
local scales.

Higher-order theories of gravity (both in metric \cite%
{review,book,capozcurv,sante,MetricRn} and Palatini \cite%
{PalRn,lnR,Allemandi} formulations) represent an interesting approach able
to fruitfully cope with both dark matter and dark energy problems. A further
approach is based on scalar\thinspace -\thinspace tensor theories of gravity
but it can be shown that higher-order theories and scalar tensor ones can be
related by conformal transformations (see, e.g., \cite{book,faraoni} and
references therein).

It has been widely demonstrated that such theories can agree with the
cosmological observations of the Hubble flow \cite{noiijmpd,noifranca} and
the large scale structure evolution \cite{frlss}. In addition, in the weak
field limit, the gravitational potential turns out to be modified \cite%
{stelle,schmidt,mannheim,noipla} in such a way that interesting consequences
to galactic dynamics may be achieved without violating, at the same time,
the constraints on the parametrized post-Newtonian (PPN) parameters coming
from Solar System tests \cite{grgOdi}.

If alternative theories of gravitation are able to explain both cosmological
and local observations without the introduction of exotic energy-momentum
sources, then one might ask how would the differences between these
alternative theories be compared to GR with the unusual sources. It is
proposed that gravitational lensing might be able to act a means for
determining which theory governs the gravitational interaction, through
measurements of image separations and the brightnesses of those multiple
images.

It is well known that the deflection of light observed during the Solar
eclipse of 1919 was one of the first experimental confirmations of
Einsteinian GR. \textit{Gravitational lensing}, i.e. the deflection of light
rays crossing the gravitational field of a compact object referred to as the 
\textit{lens}, has become one of the most astonishing successes of GR and it
represents nowadays a powerful tool capable of putting constraints on the
dynamics of gravitational structures at different scales, from stars to
galaxies and clusters of galaxies, and from the large scale structure to
cosmological parameters \cite{SEF,petters}. If we modify the Lagrangian of
the gravitational field, it is obvious that gravitational lensing should be
affected. It is therefore mandatory to investigate how gravitational lensing
behaves in the framework of alternative theories of gravity to develop a
further check on the existence of DM and DE.

In particular, one has to verify that the phenomenology of standard
gravitational lensing is recovered in the limit as the modified theory of
gravity reduces to GR, since several observations point to the validity of
GR. However, it is worth stressing that the presence of DM has to be invoked
in such cases, in particular for large-scale structure (see e.g. the case of
Bullet Cluster \cite{bullet}). On the other hand, it is worth exploring
whether deviations from the classical results for the main lensing
quantities could be detected and act as clear signatures for modified
theories of gravity. A preliminary study in this direction is in \cite{stab1}
where the gravitational lensing, in the Newtonian limit and in the Jordan
frame, for a generic $f(R,R_{\mu \nu }^{\mu \nu },R_{\mu \nu \alpha \beta
}R^{\mu \nu \alpha \beta })$ is considered. In this paper, the modifications
of the Hilbert-Einstein action are induced by corrections to the Newtonian
potential due to the Riemann tensor.

As a first step towards such an ambitious task we consider here
power\thinspace -\thinspace law fourth order theories, i.e. we replace the
Ricci scalar $R$ in the gravity Lagrangian with the function $f(R)\propto
R^{n}$ (with $n$ a positive integer) and investigate how this affects the
gravitational lensing in the theory. In Ref. \cite{LTNJC} the gravitational
lensing in $f\left( R\right) $ theories with a Yukawa-type correction in the
potential was investigated and the conclusion reached was that weak lensing
could not discern between these theories and general relativity. We revisit
this issue by using better lens equations together with a post-Newtonian
approximation, both of which can account for higher order deviations away
from general relativity. We discuss weak lensing due to compact objects of
the $R^{n}$ theory, which have been already used to analyze rotation curves
of galaxies in \cite{FRdark} and in \cite{stab2}.

The paper is organized as follows. The next section provides a short summary
of $f(R)$ theories, focusing on the special $R^{n}$ case and we review the
post-Newtonian solution representing the compact object. The modifications
from GR can be interpreted as an effective energy-momentum tensor. We
discuss the referring energy conditions in the Appendix. We discuss in
Section \ref{f} how the predictions of weak gravitational lensing are
different in the fourth order theory and in general relativity. For this we
determine the image locations, Einstein ring radii, magnification factors
and the flux ratio, for various model parameters. We then demonstrate that
the flux ratio as function of image separation has a different power-law
dependence for each model parameter. We summarize our findings in the
Concluding Remarks.

\section{A spherically symmetric compact object in $f(R)$-gravity}

\subsection{Generalities}

Let us consider now a particular class of higher order theories of gravity,
the $f(R)$-gravity theories, that are the most natural extension of GR where
the Hilbert-Einstein action is modified with a general function of the
scalar curvature \cite{review,book}. A general action is 
\begin{equation}
{\mathcal{A}}=\int {d^{4}x\sqrt{-g}\left[ f(R)+{\mathcal{L}}_{m}\right] }
\label{f(R)action}
\end{equation}%
where $f(R)$ is a generic function of the Ricci curvature scalar $R$,
differentiable at least up to the second order, $g$ is the determinant of
the metric and ${\mathcal{L}}_{m}$ is the standard matter Lagrangian.%
\footnote{%
It is possible to take into account also the Palatini approach in which the
metric $g$ and the connection $\Gamma $ are considered independent fields
(see for example \cite{ferraris}). Here we consider the Levi-Civita
connection and use the metric approach. See \cite{review, darkmetric} for a
detailed comparison between the two pictures.} The general analysis of
ghosts of f(R) gravity can be found in Ref. \cite{Biswas}. Varying the
action with respect to the metric components $g_{\mu \nu }$, one obtains the
generalized field equations that can be recast as \cite{capozcurv,FRdefl}

\begin{equation}
G_{\mu \nu }=\displaystyle{\frac{1}{f^{\prime }(R)}}\displaystyle{\Bigg \{%
\frac{1}{2}g_{\mu \nu }\left[ f(R)-Rf^{\prime }(R)\right] +f^{\prime
}(R)_{;\mu \nu }}-\displaystyle{g_{\mu \nu }\Box {f^{\prime }(R)}\Bigg \}}+%
\displaystyle{\frac{T_{\mu \nu }^{(m)}}{f^{\prime }(R)}}  \label{eq:f-var2}
\end{equation}%
where ${G_{\mu \nu }=R_{\mu \nu }-\frac{R}{2}\,g_{\mu \nu }}$ and $T_{\mu
\nu }^{(m)}$ are the Einstein tensor and the standard matter
stress\thinspace -\thinspace energy tensor, respectively. The prime denotes
derivative with respect to $R$. The two terms ${f^{\prime }(R)}_{;\mu \nu }$
and $\Box {f^{\prime }(R)}$ imply fourth order derivatives of the metric $%
g_{\mu \nu }$ so that these models are also referred to as \textit{fourth
order gravity}. Starting from Eq. (\ref{eq:f-var2}) and adopting the
Robertson\thinspace -\thinspace Walker metric, it is possible to show that
the Friedmann equations may still be written in the usual form provided that
an \textit{effective curvature fluid} (hence the name of \textit{curvature
quintessence}) is added to the matter term with energy density and pressure
depending on the choice of $f(R)$ \cite{capozcurv}.

\subsection{Power-law models with spherical symmetry}

In the absence of matter, when spherical symmetry holds, the trace of the
field equations (\ref{eq:f-var2}),%
\begin{equation}
3\Box {f^{\prime }(R)}+Rf^{\prime }(R)-2f(R)=0\ ,
\end{equation}%
combined with the $00$\thinspace -\thinspace component leads to \cite{book}%
\begin{equation}
f^{\prime }(R)\left( 3\frac{R_{00}}{g_{00}}-R\right) +\frac{1}{2}f(R)-3\frac{%
\nabla _{0}\nabla _{0}f^{\prime }(R)}{g_{00}}=0\,.  \label{master-low}
\end{equation}%
Equation (\ref{master-low}) is completely general and holds for any function 
$f(R)$. Note that even if the metric is stationary so that $\partial
_{0}g_{\mu \nu }=0$, due to the existence of non-zero Christoffel symbols
entering the second covariant derivative, $f^{\prime }(R)_{;00}=\frac{1}{2}%
g^{ij}\partial _{i}g_{00}\partial _{j}f^{\prime }(R)$\ is non-vanishing.

The simplest choice for $f(R)$ is a power-law like $f\left( R\right) \propto
R^{n}$ with $n$ the slope of the Lagrangian (clearly, with $n=1$, we recover
the Einstein theory).

In a spherically symmetric setup whenever $R$ is vanishing, constant, or
depending only on the radial coordinate $r$ (hence the space-time is
stationary), the Jebsen-Birkhoff theorem holds, and the space-time is also
static \cite{book}, \cite{new}. Therefore we search for spherically
symmetric solutions of Eqs. (\ref{eq:f-var2}). In general, we can write the
space-time metric as\thinspace :%
\begin{equation}
ds^{2}=-A(r)dt^{2}+B(r)dr^{2}+r^{2}d\Omega ^{2},  \label{eq: schwartz}
\end{equation}%
where $d\Omega ^{2}=d\theta ^{2}+\sin ^{2}{\theta }d\varphi ^{2}$ is the
line element on the unit sphere. Eq. (\ref{master-low}) reduces to%
\begin{equation}
R_{00}(r)=\frac{2n-1}{6n}\ A(r)R(r)-\frac{n-1}{2B(r)}\frac{dA(r)}{dr}\frac{%
d\ln \left\vert {R(r)}\right\vert }{dr}\ ,  \label{master-pla}
\end{equation}%
while the trace equation reads\thinspace :

\begin{equation}
\Box {R^{n-1}(r)}=\frac{2-n}{3n}R^{n}(r)\ .  \label{eq:
tracebis}
\end{equation}

Note that as soon as $n=1$, Eq. (\ref{eq: tracebis}) reduces to $R=0$,
which, when inserted into Eq. (\ref{master-pla}), gives $R_{00}=0$ and then
the standard Schwarzschild solution is recovered. In general, expressing $%
R_{00}$ and $R$ in terms of the metric (\ref{eq: schwartz}), Eqs. (\ref%
{master-pla}) and (\ref{eq: tracebis}) become a system of two nonlinear
coupled differential equations for the two functions $A(r)$ and $B(r)$.

\subsection{A low-energy, far-field solution}

We present here a solution of the low-energy approximation of the field
equations, following Ref. \cite{FRdark}.

A physically motivated hypothesis is to search for (Schwarzschild-like)
solutions of the form

\begin{equation}
A(r)=\frac{1}{B(r)}=1+\frac{2\Phi (r)}{c^{2}}~,  \label{eq:avsphi}
\end{equation}%
where $\Phi (r)$ is the gravitational potential at the distance $r$ from a
pointlike mass $m$. Further, deviations from the Newtonian $1/r$ potential
are introduced as 
\begin{equation}
\Phi \left( r;\sigma ,r_{c}\right) =-\frac{Gm}{2r}\left[ 1+\left( \frac{r}{%
r_{c}}\right) ^{\sigma }\right] ~,  \label{potential}
\end{equation}
with a strength parameter $\sigma $ and a characteristic distance $r_{c}$.
It is straightforward to see that, for $\sigma =0$, the Newtonian potential
is recovered and the metric reduces to the Schwarzschild one (GR-case). The
cases $\sigma >1$ (as the correction to the Newtonian potential
asymptotically diverges) and $\sigma =1$ (as the correction is a constant,
obstructing\ asymptotic flatness) are excluded.

With $\Phi (r)$ inserted into Eqs. (\ref{master-pla}) and (\ref{eq: tracebis}%
), they both reduce to algebraic equations, which lead to relations between $%
\sigma $ and $n$:%
\begin{equation}
(n-1)(\sigma -3)\left[ -\sigma (1+\sigma )V_{1}\eta ^{\sigma -3}\right]
^{n-1}\left[ {\mathcal{P}}_{1}+\frac{\sigma }{\eta }V_{1}{\mathcal{P}}_{0}%
\right] =0~.  \label{eq: singleeq}
\end{equation}%
Here $\eta =r/r_{c}$, $V_{1}=Gm/c^{2}r_{c}$ and%
\begin{equation}
{\mathcal{P}}_{0}=3(\sigma -3)^{2}n^{3}-(5\sigma ^{2}-31\sigma
+48)n^{2}-(3\sigma ^{2}-16\sigma +17)n-(\sigma ^{2}-4\sigma -5)\ ,
\label{eq: pz}
\end{equation}%
\begin{equation}
{\mathcal{P}}_{1}=3(\sigma -3)^{2}(1-\sigma )n^{3}+(\sigma -3)^{2}(5\sigma
-7)n^{2}-(3\sigma ^{3}-17\sigma ^{2}+34\sigma -36)n+(\sigma ^{2}-3\sigma
-4)\sigma \ .  \label{eq: pu}
\end{equation}%
Eq. (\ref{eq: singleeq}) is identically satisfied for particular values of $%
n $ and $\sigma $. It should be noted that when deriving Eq. (\ref%
{master-pla}) from Eq. (\ref{master-low}), we assumed $R\neq 0$, which
eliminates the case $n=1$. Secondly, $\sigma =3$ may also be rejected since
as noted above, all $\sigma >1$ lead to divergences. The third factor cannot
vanish as it would imply either $\sigma =0$ or $\sigma =-1$ for which $n$ is
less than unity. Then we have to solve for the last factor. For large
scale-lengths $r_{c}$ the parameter $V_{1}\ll 1$ and acts as a
post-Newtonian parameter, while both $n$ and $\sigma $ are of order unity,
hence ${\mathcal{P}}_{0}$ and ${\mathcal{P}}_{1}$ are comparable. Therefore
in the post-Newtonian regime we can approximate the last factor with%
\begin{equation}
{\mathcal{P}}_{1}(n,\sigma )\eta =0\ ,  \label{eq: singlepol}
\end{equation}%
which is an algebraic equation for $\sigma $ as function of $n$, allowing
for the following three solutions\thinspace :

\begin{equation}
\sigma =\left\{ 
\begin{array}{l}
{\frac{3n-4}{n-1}} \\ 
{\frac{12n^{2}-7n-1-\sqrt{p(n)}}{q(n)}} \\ 
{\frac{12n^{2}-7n-1+\sqrt{p(n)}}{q(n)}}%
\end{array}%
\right.  \label{eq: bn}
\end{equation}%
with\thinspace :

\begin{equation*}
p(n) = 36n^4 + 12n^3 - 83n^2 + 50n + 1 \ ,
\end{equation*}
and

\begin{equation*}
q(n)=6n^{2}-4n+2\ .
\end{equation*}%
It is easy check that, for $n=1$ the second solution (\ref{eq: bn}) gives $%
\sigma =0$, which reduces to the Newtonian potential. The other two
solutions do not fulfill this limit, hence we keep the second solution
obtaining the low-energy, far-field expression%
\begin{equation}
\sigma =\frac{12n^{2}-7n-1-\sqrt{36n^{4}+12n^{3}-83n^{2}+50n+1}}{6n^{2}-4n+2}%
\ .  \label{sigma}
\end{equation}

The dependence of $\sigma $ upon the exponent $n$ is represented in Fig. \ref%
{sigman}.

\begin{figure}[th]
\begin{center}
\includegraphics[width=6.4cm]{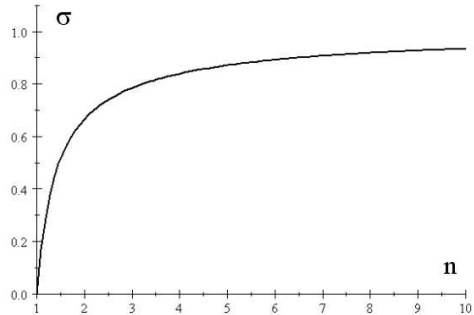}
\end{center}
\caption{The dependence of correction parameter $\protect\sigma $ in the
potential upon the exponent $n$.}
\label{sigman}
\end{figure}
In the case $n=1$ (implying $\sigma =0$) the potential reduces to the
Newtonian $\Phi _{N}$, as expected. The potential also reduces to the
Newtonian value at $r=r_{c}$. For smaller values of $r$ gravity is weakened
compared to the Newtonian values, while for $r>r_{c}$ gravity is
strengthened.

\subsection{Interpretation of the parameters}

While the power $\sigma $ of the correction term is a universal quantity
(since it depends on the exponent $n$ entering the gravity Lagrangian), the
scale-length $r_{c}$ is related to the integration constants that have to be
set to solve the fourth order differential equations of the theory. The
radius $r_{c}$ can be considered as a further gravitational radius
complementing the Schwarzschild radius, originating in the fact that we
consider a fourth order theory (compared to the second order GR) and as such
it introduces two further degrees of freedom of the gravitational field. We
expect $r_{c}$ to be related to the peculiarities of each gravitational
system. Therefore it can take different values depending upon the system's
mass and typical length scale.

\begin{figure}[th]
\begin{center}
\includegraphics[width=6.4cm]{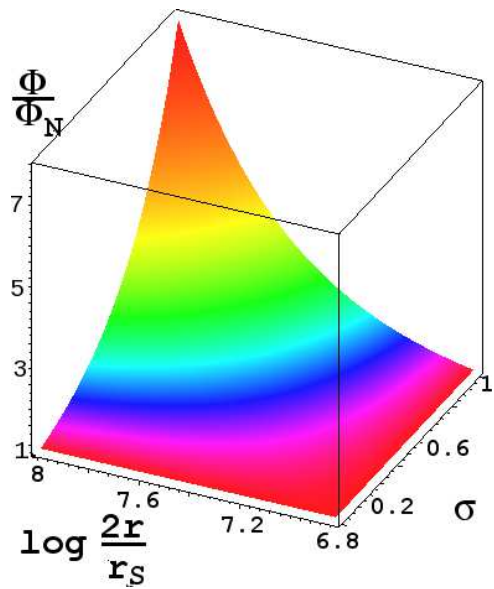}
\end{center}
\caption{The ratio of the power-law $f(R)$ gravitational potential and the
Newtonian gravitational potential as a function of $\protect\sigma \in
\lbrack 0,1)$ and of the logarithm of the distance $r\geq r_{c}$, normalized
to $r_{S}$ (the Schwarzschild radius of the lens). We have chosen a typical
bulge mass of $m_{bulge}\approx 10^{10}M_{\odot }$ and typical bulge radius $%
r_{c}\approx r_{bulge}\approx 3240$ pc.}
\label{fig1}
\end{figure}

The fact that gravity is strengthened above $r_{c}$ is illustrated in Fig. %
\ref{fig1} by plotting the ratio of the potentials $\Phi \left( r;\sigma
,r_{c}\right) /\Phi _{N}\left( r\right) $ as function of $\sigma $ and $r$
for the ranges $0\leq \sigma <1$ and $r\geq r_{c}$ with a suitably chosen
value of $r_{c}$. In the case of a typical spiral galaxy we identify $r_{c}$
with the bulge radius $r_{bulge}$, in order to have the Newtonian potential
at $r\approx r_{bulge}$, cf. Eq. (\ref{potential}). We have chosen a typical
bulge radius 3240 parsec (pc), corresponding to the order $10^{20}$\ m \cite%
{DwornikGalaxy}, and typical bulge mass of $10^{10}M_{\odot }$. Therefore
gravity is strengthened outside $r_{bulge}$ as compared to the Newtonian
case, providing an alternative to dark matter as a source for a flat
rotation curve \cite{Salucci}. We note also that the ratio $r_{c}/r_{S}$\ is
of order 10$^{6}.$

The energy conditions for the effective energy momentum tensor are
investigated in the Appendix.

\section{Weak lensing in fourth order gravity \label{f}}

\subsection{The deflection angle}

In Ref. \cite{FRdefl} the weak lensing by point-like massive objects
characterized by the potential (\ref{potential}) was investigated, where a\
deflection angle\textbf{\ }%
\begin{equation}
\delta =\frac{2Gm}{c^{2}b}\left[ 1+\frac{\sqrt{\pi }(1-\sigma )\Gamma
(1-\sigma /2)}{2\Gamma (3/2-\sigma /2)}\left( \frac{b}{r_{c}}\right)
^{\sigma }\right] ~  \label{deflection.angle}
\end{equation}%
was derived. Here $b=\left\vert \theta \right\vert D_{l}$ is the impact
parameter (defined as the distance of the lensing object to the straight
line trajectory, which would occur in the absence of the lensing object).
For $\sigma =0$ the deflection angle reproduces the Schwarzschild value $%
\delta _{S}=4Gm/c^{2}b$, while in the limit $\sigma \rightarrow 1$ the
deflection angle is one half of $\delta _{S}$.

\begin{figure}[th]
\begin{center}
\includegraphics[width=7.5cm]{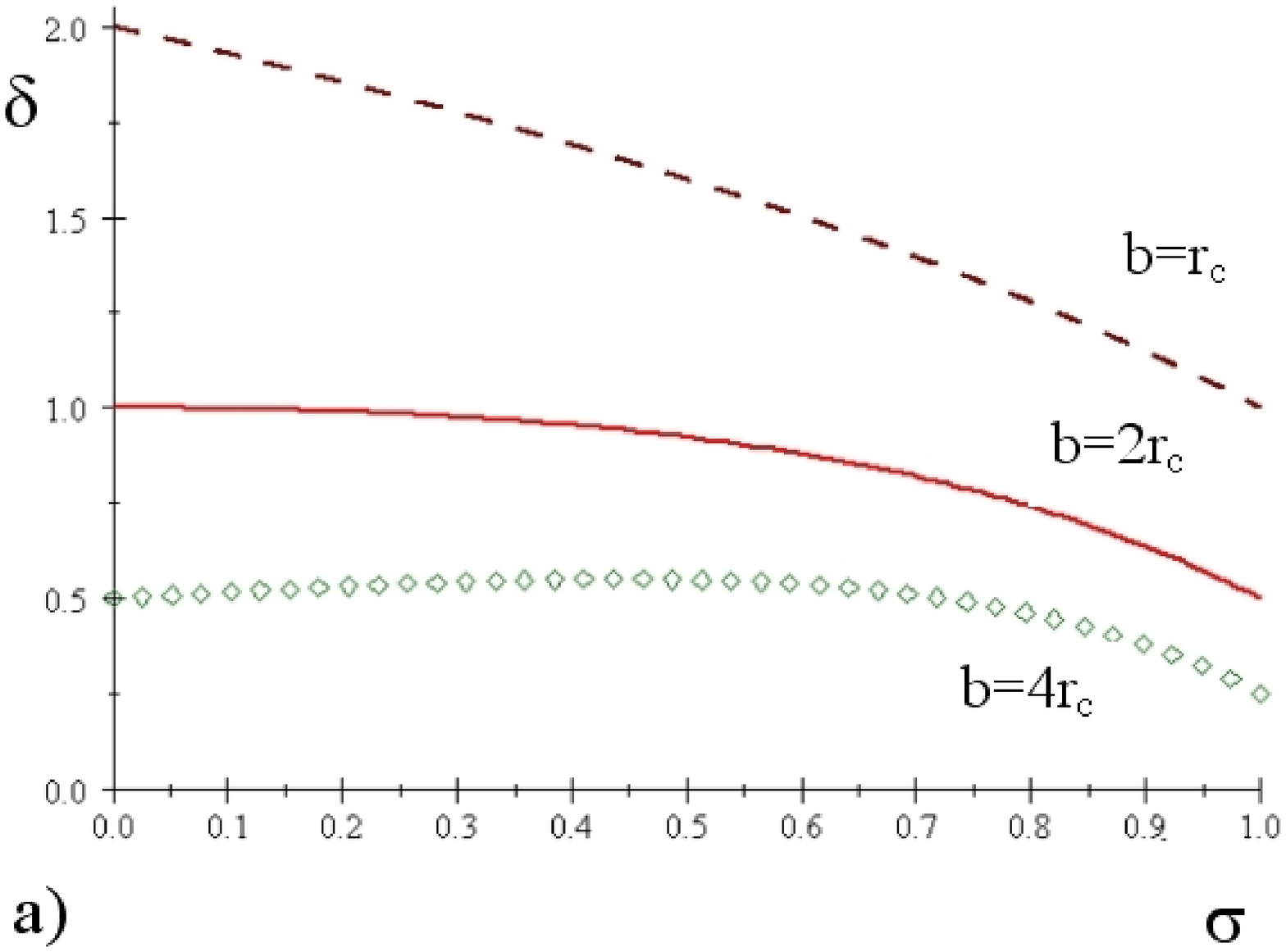} %
\includegraphics[width=7.5cm]{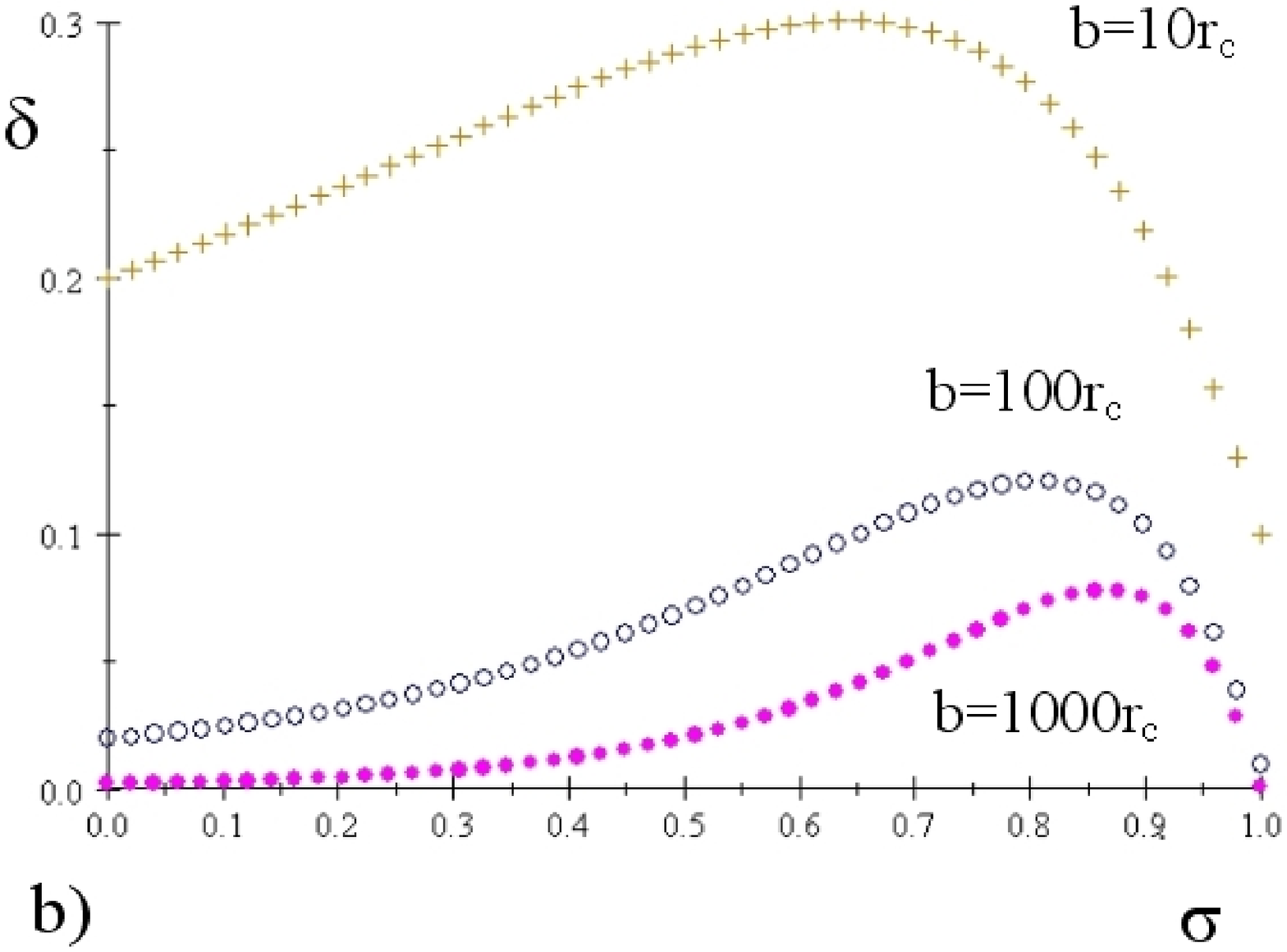}
\end{center}
\caption{The plots show the $\protect\delta (\protect\sigma )$\ dependence
in radians for different values of $b/r_{c}$, in units $2Gm/c^{2}r_{c}=1$,
with the choice $r_{c}=3240$\ pc. The respective values of $b/r_{c}$\ are
from top to bottom $1$\ (dashed)$,~2$\ (solid)$,~4$\ (diamond) on panel (a)
and $10$\ (cross)$,~100$\ (circle)$,~1000$\ (dotted) on panel (b). The
critical behavior appears at $b/r_{c}=2$.}
\label{delta}
\end{figure}

In order to investigate the behavior of $\delta $ in between the limiting
values, we represent the deflection angle as function of $\sigma $ in Fig. %
\ref{delta}. Three conclusions stand out: i) by increasing the impact
parameter at any fixed value of $\sigma $, the deflection angle always
decreases, as in the Schwarzschild case; ii) there is a critical value of
the ratio $b/r_{c}$ at $\left( b/r_{c}\right) _{crit}=2$, below which the
deflection angle monotonically decreases with increasing $\sigma $, and
above which there is a single maximum at some $\sigma _{\delta _{\max }}$;
iii) the parameter value $\sigma _{\delta _{\max }}$ increases with the
value of $b/r_{c}$. The rate of decrease of $\delta $ with increasing $b$ is
lessened as compared to the Schwarzschild case for small impact parameters
of order $r_{c}$.

\subsection{The lensing geometry}

\begin{figure}[th]
\begin{center}
\includegraphics[height=4.5cm]{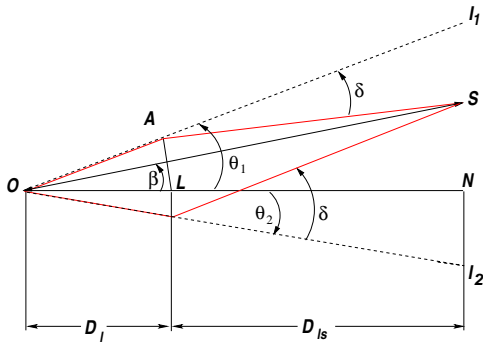}
\end{center}
\caption{The lensing geometry (see text for discussion of the symbols).
Positive angles are represented with counterclockwise directed arcs. }
\label{fig2}
\end{figure}

The lensing geometry is shown on Fig. \ref{fig2}. The optical axis $OLN$ is
defined by the observer position $O$, the lens position $L\,$, and
intersects the source plane at $N$. In the source plane $S$ represents the
location of the source and $I_{1,2}$ the locations of the two images. We use
the notations $D_{ls}$ and $D_{l}$ for the projection of the lens-source and
observer-lens distances onto the optical axis respectively \cite{TidalLens2}%
. The observer-source distance is $D_{s}=D_{l}+D_{ls}$. The source is
located at an angle $\beta $ from the optical axis, chosen to be
\textquotedblleft above" the optical axis. Images are located at angles $%
\theta _{1,2}$ with respect to the optical axis and they can be either
positive (for the image above the optical axis) or negative (for the image
below the optical axis). For either of the images we denote $s=$sgn $\theta $%
, such that $\left\vert \theta \right\vert =s\theta $. We follow the
convention that the deflection angle is $\delta >0$ whenever the light is
bent towards the optical axis, cf. Ref. \cite{Virbhadra}. Similarly as in
Ref. \cite{TidalLens2}, we characterize the mass by the dimensionless
parameter $\bar{\varepsilon}=Gm/c^{2}L$, with $L=D_{s}D_{l}/D_{ls}$.

In Ref. \cite{FRdefl} the leading order lens equation 
\begin{equation}
\left\vert \theta \right\vert -s\beta -\frac{D_{ls}}{D_{s}}\delta =0
\label{lens0}
\end{equation}%
was employed for the discussion of the weak lensing. Many authors have
obtained more accurate results for weak lensing by using explicit
trigonometric relationships. One of the most useful is that given by
Virbhadra and Ellis \cite{VE}, where their lens equation is 
\begin{equation}
\tan \left\vert \theta \right\vert -\tan \left( s\beta \right) -\frac{D_{ls}%
}{D_{s}}\left[ \tan \left\vert \theta \right\vert +\tan \left( \delta
-\left\vert \theta \right\vert \right) \right] =0~.  \label{lens_Ellis}
\end{equation}%
Alternatively even more general lens equations have been derived in Ref. 
\cite{DabrowskiSchunck}: 
\begin{equation}
\sin \alpha =\frac{D_{ls}}{D_{s}}\cos \left\vert \theta \right\vert \cos %
\left[ \arcsin \left( \frac{D_{s}}{D_{ls}}\sin \left( \left\vert \theta
\right\vert -\alpha \right) \right) \right] \left[ \tan \left\vert \theta
\right\vert +\tan \left( \delta -\left\vert \theta \right\vert \right) %
\right] ~,  \label{ds}
\end{equation}%
where $\alpha :=\theta -\beta $ is the "reduced" deflection angle, or in
Ref. \cite{TidalLens2}:%
\begin{equation}
\frac{2D_{l}}{D_{s}}\cos \left( \frac{\delta }{2}-\left\vert \theta
\right\vert \right) \cos \left\vert \theta \right\vert \sin \frac{\delta }{2}%
+\cos \left( \delta -\left\vert \theta \right\vert \right) \left( \sin
\left\vert \theta \right\vert -s\cos \left\vert \theta \right\vert \tan
\beta \right) -\sin \delta ~=0.  \label{tidalle}
\end{equation}

Which expression to use in the subsequent calculations will depend on the
ability of current technology to resolve the differences between the
predictions made by the various proposed lens equations. Assuming that the
observations of the lensed image positions are (or soon will be) capable of
resolving angular differences on the order of microarcseconds (see e.g. Ref. 
\cite{Jin2008}) one can compare the accuracy of the different lens equations
and then choose the simplest expression that is compatible with the
available astrometric precision.

\begin{figure}[th]
\begin{center}
\includegraphics[height=6.5cm]{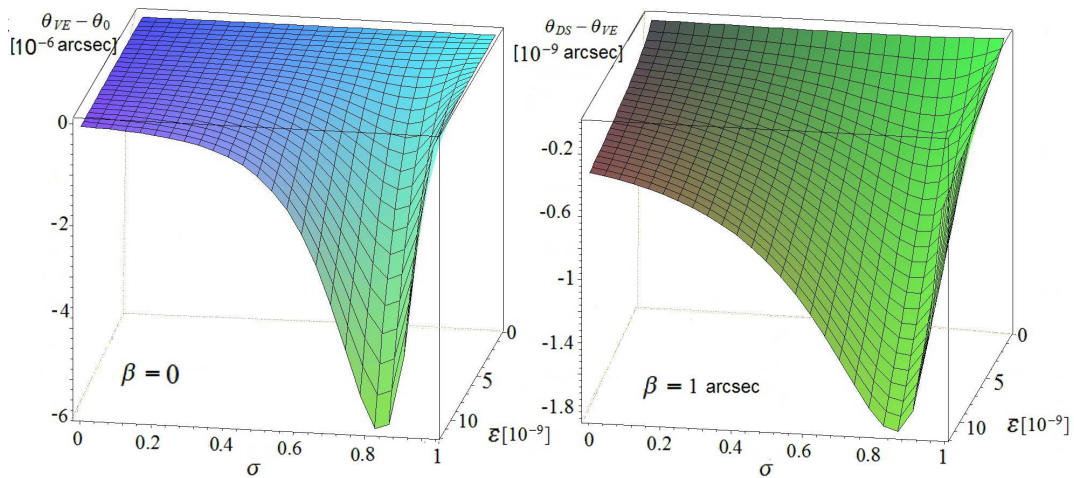}
\end{center}
\caption{(a) The difference between the Einstein angles $\protect\theta %
_{VE} $, obtained from Eq. (\protect\ref{lens_Ellis}) and $\protect\theta %
_{0}$\ , obtained from Eq. (\protect\ref{lens0}), as function of the
parameters $\protect\sigma $\ and $\bar{\protect\varepsilon}$\ . (b) The
difference between the positive apparent angles $\protect\theta _{VE}$\ and $%
\protect\theta _{DS}$\ [obtained from Eq. (\protect\ref{ds})] of the images
of a source located at $\protect\beta =1$ arcsec. While the two surfaces
have the same shape, it should be noted that the vertical scales differ by 3
orders of magnitude. We set the distances $D_{l}=1$\ Mpc and $D_{ls}=2$\
Mpc. }
\label{compareangle}
\end{figure}

The left panel of Fig. \ref{compareangle}. compares the lens equations (\ref%
{lens_Ellis}) and (\ref{lens0}), while right panel compares the lens
equations (\ref{ds}) and (\ref{lens_Ellis}). The difference between the
Einstein angles $\theta _{VE}$, obtained from Eq. (\ref{lens_Ellis}) and $%
\theta _{0}$, obtained from Eq. (\ref{lens0}), are represented as function
of the parameters $\sigma $\ and $\bar{\varepsilon}$\ on the left panel of
Fig. \ref{compareangle}, and they are of order of microarcseconds. The
differences between the Einstein angles $\theta _{VE}$\ and $\theta
_{DS}^{{}}$, obtained from Eq. (\ref{ds}) are shown on the right panel of
Fig. \ref{compareangle}, and they are of order of nanoarcseconds. Remarkably
the two figures are identical in shape, but differing in 3 orders of
magnitude. A comparison between Eqs. (\ref{tidalle}) and (\ref{lens_Ellis})
leads to a similar conclusion.

Hence Fig. \ref{compareangle}. illustrates that both Eqs. (\ref{lens_Ellis})
and (\ref{ds}) lead to identical corrections of Eq. (\ref{lens0}) in the
lensing behavior within less than 1\%. In what follows we employ the simpler
lens equation (\ref{lens_Ellis}), rather than (\ref{ds}) or the second order
lens equation derived in Ref. \cite{TidalLens2}, as it provides sufficient
accuracy for confrontation with observations.

\subsection{Image positions and magnifications}

In Ref. \cite{FRdefl} the image positions, the radius of the Einstein ring,
image magnifications and the Paczynski curve in microlensing experiments
were also estimated. In this follow-up paper we improve on the accuracy of
the weak lensing characteristics, by employing a more accurate lens equation
with the focus on the behavior of dimensionless observable quantities which
can be derived from the image positions and magnification factors. Our task
is to determine the measurable differences between the predictions of the
fourth order theory and those of GR. 
\begin{figure}[t]
\begin{center}
\includegraphics[height=7.5cm]{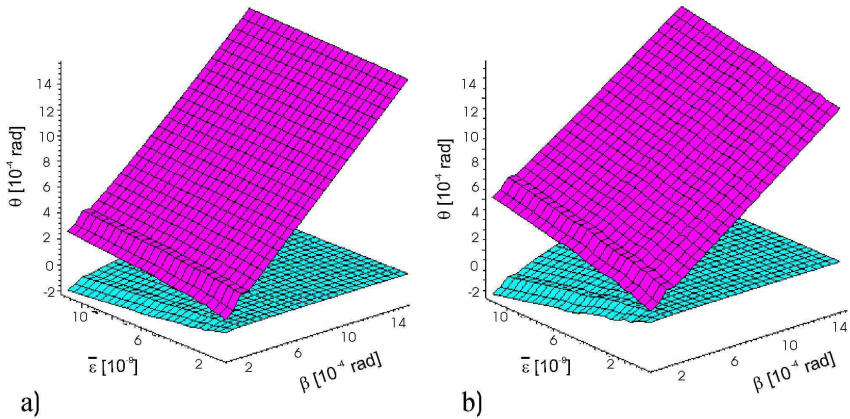}
\end{center}
\caption{The image positions $\protect\theta $ as function of $\bar{\protect%
\varepsilon}$ and $\protect\beta $ for $\protect\sigma =0.25$ (a) and $%
\protect\sigma =0.75$ (b), for the distances $D_{l}=1$ Mpc and $D_{ls}=2$
Mpc. The angle $\protect\beta $ is varied up to $0.0015$ rad, similarly as
on Fig 4(b) of Ref. \protect\cite{TidalLens2}. With decreasing $\protect%
\beta $, the image separations shrink accordingly. At $\protect\beta =0$ the
angle $\protect\theta $ represents the angular radius of the Einstein ring.
As we expect the $\protect\beta =0$ sections of the surfaces are symmetric
with respect to the plane $\protect\theta =0$.}
\label{images}
\end{figure}

A weak lens equation was derived in Ref. \cite{TidalLens2} exclusively by
trigonometric considerations and was applied to the computations of the
image positions, magnifications and flux ratios to second order accuracy
(both in the mass-related and tidal charge related small parameters) for
brane-world black holes. In addition it was shown, how the Virbhadra-Ellis
lens equation follows as an approximation (agreement is reached in the first
order of the perturbations). For the purposes of the present paper, it is
recognized that second order and higher effects will not be measurable by
current telescope technology, and therefore we will utilize the
Virbhadra-Ellis lens equation \cite{Virbhadra}, \cite{VE}, \cite{Jin2008}, 
\cite{Bozza}, together with the deflection angle (\ref{deflection.angle})
derived in Ref. \cite{FRdefl}. This generalizes the approach of Ref. \cite%
{FRdefl}, where the leading order lens equation [Eq. (12) there] was
employed for the discussion of weak lensing effects.

The numerical solution of the system of equations (\ref{deflection.angle})
and (\ref{lens_Ellis}) gives the positions of the images as function of $%
\bar{\varepsilon}$ and $\beta $, represented on Fig. \ref{images}, for $%
\sigma =0.25$ and $0.75$, respectively. In both cases decreasing $\beta $
decreases the image separations. As expected, the $\beta =0$ sections give
symmetric curves with respect to the planes $\theta =0.$ This is because at $%
\beta =0$ the angle $\theta =\theta _{E}$ represents the angular radius of
the Einstein ring. For small impact parameter (implying small $\beta $) the
image separations obey $\left( \theta _{1}-\theta _{2}\right) _{\sigma
=0.25}<\left( \theta _{1}-\theta _{2}\right) _{\sigma =0.75}$, whereas for
large values of $\beta $ the image separations behave as $\left( \theta
_{1}-\theta _{2}\right) _{\sigma =0.25}>\left( \theta _{1}-\theta
_{2}\right) _{\sigma =0.75}$. This indicates that our analysis based on the
Virbhadra-Ellis lens equation is more accurate than the first post-Newtonian
order calculation performed in Section VIII. in Ref. \cite{bg}, which states
that $f(R)$ gravity is indistinguishable from general relativity and is
consistent with the observational value of the post-Newtonian parameter $%
\gamma =1+(2.1\pm 2.3)\times 10^{-5}$. 
\begin{figure}[th]
\begin{center}
\includegraphics[height=6.5cm]{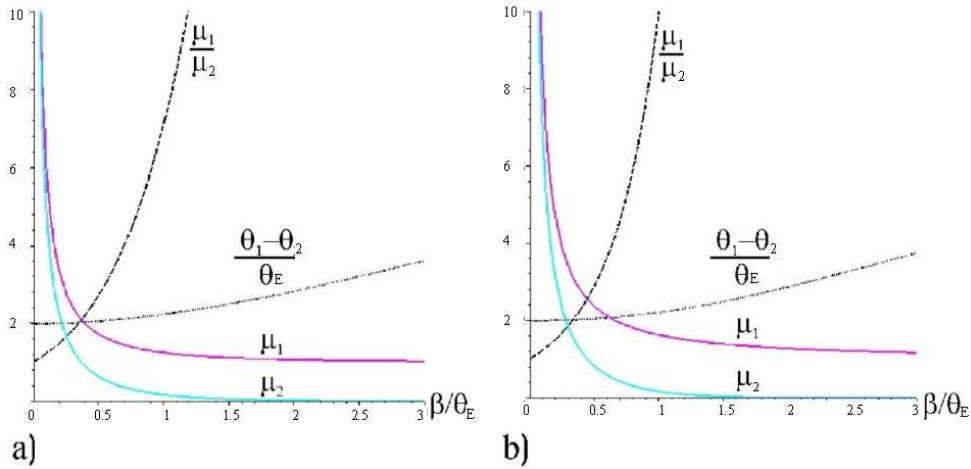}
\end{center}
\caption{The image separations and magnifications as functions of $\protect%
\beta /\protect\theta _{E}$ for $\protect\sigma =0.25$ (a) and $\protect%
\sigma =0.75$ (b). We fixed $\bar{\protect\varepsilon}=3.375\times 10^{-9}$,
while the distances $D_{l}=1$ Mpc and $D_{ls}=2$ Mpc were chosen for the
plots. The upper and lower solid curves represent the primary and secondary
image magnification factors, respectively; their ratio is the dashed curve;
and the dotted curve is the normalized image separation.}
\label{magsep}
\end{figure}

The magnification of the images are defined as 
\begin{equation}
\mu _{1,2}=\left\vert \frac{d\theta _{1,2}}{d\beta }~\frac{\theta _{1,2}}{%
\beta }\right\vert ~.
\end{equation}%
For Schwarzschild lensing $\mu _{1}-\mu _{2}=1$ always holds. Figure \ref%
{magsep} shows the image separations and magnifications as functions of $%
\beta /\theta _{E}$ for $\sigma =0.25$ and $0.75$. The upper and lower solid
curves represent the primary and secondary image magnification factors,
respectively; their ratio is the dashed curve; and the dotted curve is the
image separation normalized with respect to the Einstein angle. The
strongest effect appears on the ratio of magnifications, which increases
more rapidly with $\beta $ as $\sigma $ increases. The reason for this is
that the primary image is magnified more for larger values of $\sigma $.

\subsection{Power-law behavior}

In a lensing observation the most straightforward measurements are 1) the
angular separation between the two images and 2) the ratio of the
magnification factors. The first measurement does not require information on
the location of the lens position, needed to define the individual image
positions. The second measurement does not require an absolute measure of
image brightness, since we are taking a ratio between the two.

\begin{figure}[th]
\begin{center}
\includegraphics[width=13cm]{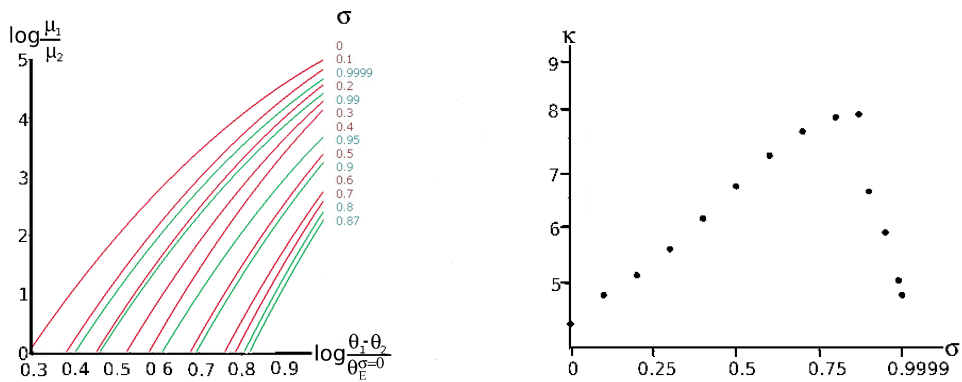} \ \ \ \ \ \ \ \ \ \ %
\end{center}
\caption{The ratio of the magnification factor of the primary and secondary
images as function of the image separation (normalized to the Einstein angle 
$\protect\theta _{E}^{\protect\sigma =0}$, characterizing the $\protect%
\sigma =0$ case), on log--log scale for a series of $\protect\sigma $ values
(left panel). There is a double degeneracy in $\protect\sigma $ of the power 
$\protect\kappa $, represented as a function of $\protect\sigma $ (right
panel). The plot refers to $\bar{\protect\varepsilon}=3.375\times 10^{-9}$
and distances $D_{l}=1$ Mpc, $D_{ls}=2$ Mpc.}
\label{loglog}
\end{figure}

Therefore we plot the ratio of the magnification factor for the primary and
secondary images as function of the image separation (normalized to the
Einstein angle), on the log--log scale, in Fig. \ref{loglog}. The $\sigma =0$
curve characterizes the lensing by a Schwarzschild black hole, the colored
curves correspond to the fourth order gravity lensing for the parameter
values of $\sigma $ ranging from 0.1 to 0.9999.

Similar to Ref. \cite{TidalLens2}, it is found that for image separations
greater than about 2.5 times the Einstein angle, the ratio of the
magnification factors for each value of $\sigma $ obeys a power-law
relationship%
\begin{equation}
\frac{\mu _{1}}{\mu _{2}}\propto \left( \frac{\Delta \theta }{\theta _{E}}%
\right) ^{\kappa }~.
\end{equation}%
The different slopes $\kappa $ of the curves indicate power-law behaviors
with different exponents, which are presented in Table \ref{table1} and in
Fig. \ref{loglog}.

\begin{table*}[t]
\caption{The exponents $\protect\kappa$ associated with the power law
scaling of the magnification ratio as function of image separation for
various values of $\protect\sigma$, up to a double degeneracy.}
\label{table1}%
\begin{tabular}{|c||c|c|c|c|c|c|c|c|c|c|c|c|c|c|}
\hline
$\sigma $ & $0$ & $0.1$ & $0.2$ & $0.3$ & $0.4$ & $0.5$ & $0.6$ & $0.7$ & $%
0.8$ & $0.87$ & $0.9$ & $0.95$ & $0.99$ & $0.9999$ \\ \hline
$\kappa $ & $4.39$ & $4.87$ & $5.21$ & $5.64$ & $6.15$ & $6.69$ & $7.20$ & $%
7.60$ & $7.84$ & $7.92$ & $6.60$ & $5.92$ & $5.12$ & $4.87$ \\ \hline
\end{tabular}%
\end{table*}

Given a large enough number of measurements of image separations and image
brightnesses, these power-law relations provide an observational signature
that can distinguish among the fourth order $f(R)$ theories with different $%
\sigma $ (or $n$).

\section{Concluding remarks}

In this paper we have analyzed the weak lensing signatures of a fourth order
[$f\left( R\right) =R^{n}$] gravity compact object, with gravitational
potential given in the post-Newtonian regime by Eqs. (\ref{potential}) and (%
\ref{sigma}). This introduces a new parameter $\sigma \in \lbrack 0,1)$,
which governs the deviation from the Newtonian gravitational potential for
different values of $n$. General relativity is contained as the special case 
$n=1$ (corresponding to the model parameter $\sigma =0$). For any other
value of the parameter $n$ (or $\sigma $) the gravitational attraction
increases at distances larger than $r_{c}$ as compared to the prediction of
Newtonian gravity.

The lensing properties of such compact objects were analyzed before in Ref. 
\cite{FRdefl}, based on the small angle lens equation \cite{Bozza}. In this
paper we have improved upon this approach, by employing the first order
accurate Virbhadra-Ellis lens equation (\ref{lens_Ellis}), or with
equivalent results the D\'{a}browski-Schunck lens equation (\ref{ds}).

We analyzed the dependences upon $\sigma $ and upon the impact parameter $b$
of the deflection angle (\ref{deflection.angle}). The deflection angle
decreases with increasing impact parameter for all $\sigma $. There is a
transition at a critical value $\left( b/r_{c}\right) _{crit}=2$, below
which the deflection angle monotonically decreases with increasing $\sigma $%
, and above which there is a single maximum. This maximum value increases
with the impact parameter.

The image positions as a function of the lensing mass and source position,
also the image magnifications and their ratio as function of source position
show features similar to those in the Schwarzschild case. Nevertheless, in
contrast with previous claims in the literature \cite{stab1}, \cite{LTNJC}
these lensing quantities depend upon $\sigma $.

We have computed the image positions for two values of $\sigma $. For the
larger value of $\sigma $, the image separation grows faster with an
increase in the mass and grows more slowly as the source moves away from the
optical axis.

For the same source position the magnification factors of the images
increase with $\sigma $, especially the one for the primary image. The
increases in their ratio $\mu _{1}/\mu _{2}$ is even more significant.

Using the most easily measurable lensing observables, the ratio of the
magnifications is shown to have a power-law dependence on the image
separations, with the power strongly depending on $\sigma $. The power is
the smallest for Schwarzschild black holes ($\sigma =0$), then it increases
with $\sigma $ to a critical value, after which it decreases again. This
behavior provides a means for future gravitational lensing observations to
either establish the value of $\sigma $ up to a double degeneracy or falsify
the power-law gravitational potential discussed in this paper if $\sigma =0$
is confirmed. Given that the next generation of radio telescopes will easily
be able to resolve images to less than milliarcsecond (in fact tens of
microarcseconds\textbf{)} accuracy, the different rates at which the ratio
of the magnifications changes should be able to provide a significant
observational signature constraining the validity of $f\left( R\right) $
gravitational theories.

\section*{Acknowledgements}

Z. H. and L. \'{A}. G. were supported by the European Union and European
Social Fund Grants T\'{A}MOP 4.2.2/B-10/1-2010-0012 and T\'{A}%
MOP-4.2.2.A-11/1/KONV-2012-0060, by COST Action MP0905 "Black Holes in a
Violent Universe", and by the Hungarian Scientific Research Fund (OTKA)
Grant No. 81364. D. H. acknowledges support from an NSERC Discovery Grant.

\appendix

\section{Energy conditions in the low-energy, far-field regime}

The low-energy, far-field solution (\ref{eq: schwartz}), (\ref{eq:avsphi})-(%
\ref{potential}) and (\ref{sigma}) holds for vacuum, hence there are no
energy conditions to observe in the $f\left( R\right) $ theory.

The first term in Eq. (\ref{eq:f-var2}) however can be interpreted as an
effective energy-momentum tensor of geometric origin within the context of
General Relativity. In terms of the metric decomposition 
\begin{equation}
g_{\mu \nu }=-u_{\mu }u_{\nu }+n_{\mu }n_{\nu }+h_{\mu \nu }~,
\end{equation}%
[with $h_{\mu \nu }$ the metric on the 2-sphere surfaces, $u^{\mu }=\left( 1/%
\sqrt{A},0,0,0\right) $ the temporal unit vector and $n^{\mu }=\left( 0,%
\sqrt{A},0,0\right) $ the radial unit vector], the energy-momentum tensor of
the spherically symmetric, static curvature fluid is decomposed as 
\begin{equation}
T_{\mu \nu }=\rho u_{\mu }u_{\nu }+p_{r}n_{\mu }n_{\nu }+p_{t}h_{\mu \nu }~,
\end{equation}%
or 
\begin{equation}
T_{\mu \nu }=\left( \rho +p_{t}\right) u_{\mu }u_{\nu }+\left(
p_{r}-p_{t}\right) n_{\mu }n_{\nu }+p_{t}g_{\mu \nu }~,
\end{equation}%
Here $\rho $ is the energy density, $p_{r}$ and $p_{t}$ the the radial and
tangential pressures, respectively: 
\begin{eqnarray}
\rho &=&T_{\mu \nu }u^{\mu }u^{\nu }=T_{00}~,  \notag \\
p_{r} &=&T_{\mu \nu }n^{\mu }n^{\nu }=T_{11}~,  \notag \\
p_{t} &=&\frac{1}{2}\left( T_{\mu \nu }h^{\mu \nu }\right) =\frac{T_{22}}{%
r^{2}}~.  \label{rhop}
\end{eqnarray}%
The equivalent energy-momentum tensor for the fourth order theory reads:%
\begin{equation}
T_{\mu \nu }^{equiv}=\left( {1-n}\right) \left\{ \frac{1}{2n}{g_{\mu \nu }}%
R+\left( g_{\mu }^{\rho }g_{\nu }^{\sigma }{-{g_{\mu \nu }g}^{\rho \sigma }}%
\right) \left[ \left( 1-n\right) \nabla _{\rho }\left( \ln \left\vert
R\right\vert \right) \nabla _{\sigma }\left( \ln \left\vert R\right\vert
\right) -\nabla _{\rho }\nabla _{\sigma }\left( \ln \left\vert R\right\vert
\right) \right] \right\} ~,
\end{equation}%
vanishing, as required, in the GR case $n=1$. Here the curvature scalar is
given by 
\begin{eqnarray}
R &=&\frac{2\left( A-1\right) }{r^{2}}+\frac{4A_{,r}}{r}+A_{,rr}  \notag \\
&=&-\sigma \left( \sigma +1\right) \frac{Gm}{c^{2}r_{c}^{\sigma }}r^{\sigma
-3}~,
\end{eqnarray}%
while 
\begin{eqnarray}
\nabla _{\sigma }\ln \left\vert R\right\vert &=&\frac{\left( \sigma
-3\right) }{r}\delta _{\sigma }^{1}~,  \notag \\
\nabla _{\rho }\nabla _{\sigma }\left( \ln \left\vert R\right\vert \right)
&=&-\frac{\left( \sigma -3\right) }{r^{2}}\delta _{\rho }^{1}\delta _{\sigma
}^{1}-\frac{\left( \sigma -3\right) }{r}\Gamma _{\rho \sigma }^{1}~,
\end{eqnarray}%
where the required Christoffel symbols are%
\begin{eqnarray}
\Gamma _{00}^{1} &=&-\Gamma _{11}^{1}=\frac{AA_{,r}}{2}=\frac{Gm}{2c^{2}r^{2}%
}\left[ 1+\left( 1-\sigma \right) \left( \frac{r}{r_{c}}\right) ^{\sigma }%
\right] \left( 1-\frac{Gm}{c^{2}r}\left[ 1+\left( \frac{r}{r_{c}}\right)
^{\sigma }\right] \right)  \notag \\
&\approx &\frac{Gm}{2c^{2}r^{2}}\left[ 1+\left( 1-\sigma \right) \left( 
\frac{r}{r_{c}}\right) ^{\sigma }\right] ~, \\
\Gamma _{22}^{1} &=&\frac{\Gamma _{33}^{1}}{\sin ^{2}\theta }=-rA=-r\left( 1-%
\frac{Gm}{c^{2}r}\left[ 1+\left( \frac{r}{r_{c}}\right) ^{\sigma }\right]
\right) ~.
\end{eqnarray}%
The $\approx $ denotes expansion to first order in the post-Newtonian
parameter $Gm/c^{2}r$.

The equivalent energy-momentum tensor becomes%
\begin{eqnarray}
T_{\mu \nu }^{equiv} &=&{\left( {1-n}\right) g_{\mu \nu }}\left\{ -{\frac{%
\sigma \left( \sigma +1\right) Gm}{2nc^{2}r_{c}^{\sigma }}r^{\sigma -3}+}%
\frac{\left( \sigma -3\right) }{r^{2}}\left[ \left[ 1+\left( 1-n\right)
\left( \sigma -3\right) \right] A+r{{g}^{\rho \sigma }}\Gamma _{\rho \sigma
}^{1}\right] \right\}  \notag \\
&&+{\left( {1-n}\right) }\frac{\left( \sigma -3\right) }{r^{2}}\left\{ \left[
\left( 1-n\right) \left( \sigma -3\right) +1\right] g_{\mu }^{1}g_{\nu
}^{1}+r\Gamma _{\mu \nu }^{1}\right\}
\end{eqnarray}%
From Eqs. (\ref{rhop}) and the intermediary result%
\begin{eqnarray}
{{g}^{\rho \sigma }}\Gamma _{\rho \sigma }^{1} &=&\frac{\left(
1-A^{2}\right) A_{,r}}{2}+2r^{-1}A  \notag \\
&=&\frac{1}{4r}\left( \frac{Gm}{c^{2}r}\right) ^{2}\left[ 1+\left( \frac{r}{%
r_{c}}\right) ^{\sigma }\right] \left[ 1+\left( 1-\sigma \right) \left( 
\frac{r}{r_{c}}\right) ^{\sigma }\right] \left( 1-\frac{Gm}{4c^{2}r}\left[
1+\left( \frac{r}{r_{c}}\right) ^{\sigma }\right] \right)  \notag \\
&&+2r^{-1}\left\{ 1-\frac{Gm}{2c^{2}r}\left[ 1+\left( \frac{r}{r_{c}}\right)
^{\sigma }\right] \right\}  \notag \\
&\approx &2r^{-1}\left\{ 1-\frac{Gm}{2c^{2}r}\left[ 1+\left( \frac{r}{r_{c}}%
\right) ^{\sigma }\right] \right\}
\end{eqnarray}%
we obtain the first order post-Newtonian expressions%
\begin{eqnarray}
\rho &=&\frac{{n-1}}{r^{2}}\left( \rho _{r}^{0}-\frac{Gm}{c^{2}r}\rho
_{r}^{1}\right) ~,  \notag \\
p_{r} &=&\frac{{1-n}}{r^{2}}\left( p_{r}^{0}-\frac{Gm}{c^{2}r}%
p_{r}^{1}\right) ~,  \notag \\
p_{t} &=&\frac{{1-n}}{r^{2}}\left( p_{t}^{0}-\frac{Gm}{c^{2}r}%
p_{t}^{1}\right) ~,
\end{eqnarray}%
with the leading order contributions%
\begin{eqnarray}
\rho _{r}^{0} &=&\frac{{n-1}}{r^{2}}\left( \sigma -3\right) \left[ 3+\left(
n-{1}\right) \left( 3-\sigma \right) \right] ~,  \notag \\
p_{r}^{0} &=&2\frac{{1-n}}{r^{2}}\left( \sigma -3\right) \left[ 2+\left( n-{1%
}\right) \left( 3-\sigma \right) \right] ~,  \notag \\
p_{t}^{0} &=&\frac{{1-n}}{r^{2}}\left( \sigma -3\right) \left[ 2+\left( n-{1}%
\right) \left( 3-\sigma \right) \right] ~,
\end{eqnarray}%
and the post-Newtonian contributions%
\begin{eqnarray}
\rho _{r}^{1} &=&2\rho _{r}^{0}+\frac{\left( \sigma -3\right) }{2}+\left(
2\rho _{r}^{0}+\rho _{c}^{1}\right) \left( \frac{r}{r_{c}}\right) ^{\sigma
}~,  \notag \\
p_{r}^{1} &=&\frac{\sigma -3}{2}+\rho _{c}^{1}\left( \frac{r}{r_{c}}\right)
^{\sigma }~,  \notag \\
p_{t}^{1} &=&p_{t}^{0}+\left[ p_{t}^{0}+{\frac{\sigma \left( \sigma
+1\right) }{2n}}\right] \left( \frac{{r}}{r_{c}}\right) ^{\sigma }~,
\end{eqnarray}%
where we have denoted 
\begin{equation}
\rho _{c}^{1}=\frac{1}{2}\left[ {\frac{\sigma \left( \sigma +1\right) }{n}}%
+\left( 1-\sigma \right) \left( \sigma -3\right) \right] ~.
\end{equation}

In what follows, we discuss various energy-conditions, which for an
spherically symmetric, static, but anisotropic fluid are subject to the
following restrictions:

\begin{itemize}
\item[i)] Weak energy conditions: $\rho \geq 0$, $\rho +p_{r}>0$ and $\rho
+p_{t}>0.$

\item[ii)] Null energy conditions: $\rho +p_{r}\geq 0$ and $\rho +p_{t}\geq
0 $.

\item[iii)] Strong energy conditions: $\rho +p_{r}\geq 0$, $\rho +p_{t}\geq
0 $ and $\rho +p_{r}+2p_{t}\geq 0$.

\item[iv)] Dominant energy conditions: $\rho \geq 0$, $\rho \geq \left\vert
p_{r}\right\vert $ and $\rho \geq \left\vert p_{t}\right\vert $.
\end{itemize}

In the weak-field approximation the leading order contributions give%
\begin{eqnarray}
\rho _{r}^{0} &=&-\frac{\left( n-{1}\right) \left( 3-\sigma \right) }{r^{2}}%
\left[ 3+\left( n-{1}\right) \left( 3-\sigma \right) \right] \leq 0~,  \notag
\\
\rho ^{0}+p_{r}^{0} &=&\frac{\left( n-{1}\right) \left( 3-\sigma \right) }{%
r^{2}}\left[ 1+\left( n-{1}\right) \left( 3-\sigma \right) \right] \geq 0~, 
\notag \\
\rho ^{0}+p_{t}^{0} &=&-\frac{\left( n-{1}\right) \left( 3-\sigma \right) }{%
r^{2}}\leq 0~,  \notag \\
\rho ^{0}+p_{r}^{0}+2p_{t}^{0} &=&\frac{\left( n-{1}\right) \left( 3-\sigma
\right) }{r^{2}}\left[ 5+3\left( n-{1}\right) \left( 3-\sigma \right) \right]
\geq 0~.
\end{eqnarray}%
The inequalities follow from the conditions $n\geq 1$ and $\sigma \in
\lbrack 0,1)$. Hence to leading order the curvature fluid violates all of
the above energy conditions, except for the vacuum GR limit $n=1$. The first
order post-Newtonian contributions cannot change this conclusion.

Based on the equivalence of metric $f\left( R\right) $\ theories with
Brans-Dicke metric-scalar field theories with $\omega =0$\ [Eqs. (3.97) and
(3.101) of Ref. \cite{book}], Ref. \cite{Sotiriou} has proven that the only
spherically symmetric, asymptotically flat vacuum black hole solution for a
scalar field with energy-momentum tensor obeying the weak energy condition
is the Schwarzschild solution.

The weak-field solution of $f(R)$ theory discussed in this paper leads to an
effective energy-momentum tensor that violates all energy conditions.
Therefore it can be considered as an approximation of a compact object (for
example a black hole) whose parameters fall outside the conditions of the
above theorem.

\end{document}